\documentclass[
  twoside,
  twocolumn,
  aps,
  preprintnumbers,
 floatfix,
  showkeys,
] {revtex4}

\usepackage[dvips]{graphicx}
\usepackage{tabularx}
\usepackage{dcolumn}
\usepackage{color}
\usepackage{amsmath}

\begin{document}
\title{Structure and magnetic properties of six-atom silver clusters supported on LTA zeolite}
\author{Amgalanbaatar Baldansuren$^{\dag}$}
\email[e-mail:\quad]{amgalanbaatar.baldansuren@manchester.ac.uk}
\altaffiliation{$^\dag$Previous address:\quad Institut f\"ur Physikalische Chemie, Universit\"at Stuttgart, D-70569 Stuttgart, Germany}
\affiliation{The Photon Science Institute, EPSRC National EPR Facility and Service, School of Chemistry, The University of Manchester, Oxford Road, Manchester M13 9PL, UK}
\date{March 29, 2015}

\begin{abstract}
12\% (wt.) silver containing catalysts were prepared by aqueous ion-exchange of Ag$^+$ against Na$^+$ in NaA zeolite. Hydrogen reduction leads to the formation of silver clusters. The results of X-ray absorption spectroscopy experiments at the Ag $K$-edge using direct {\it in-situ} reduction yield a Ag-Ag coordination number of 4.00, which is compatible with a quite mono-disperse structure of ca. six-atomic clusters. $<$ 0.1\% of them is EPR active, revealing isotropic parameters with six equivalent silver nuclei. The magnetic susceptibility in a temperature range of 1.8 K to 300 K exhibits Curie-Weiss behavior, with a magnetization up to $\approx$ 0.81 $\mu_{\rm B}$ per Ag atom, dependent on reduction history. Thus, the overall magnetization reveals many more unpaired electrons than are observed by EPR, but is still less than a single unpaired electron per atom on average. This suggests that a mixture of diamagnetic, EPR active spin-1/2 and EPR silent high-spin species is present. Interestingly, neutral free Ag$_6$ clusters were predicted by theoretical calculations to be diamagnetic and of planar triangular geometry ({\it Phys. Rev. A.} {\bf 75} (2007) 063204).
\end{abstract}

\maketitle

\section{Introduction}

Metal nano-clusters have been considered {\it strange morsels of matter} \cite{Poo90}. Depending upon the temperature and especially on the applied physical method of investigations, the clusters may still behave as a piece of metal or as a molecular system which can be described by means of quantum mechanics rather than classical physics. In the size regime of $\approx$ 2 - 5 nm, quantum properties dominate for metal clusters which indeed exhibit the discrete energy levels like atoms do \cite{Kub62,Kaw70,Hal86}. Thus, the cluster physics may occupy a unique position in condensed matter physics in such a way that physical phenomena can be more directly linked to the experimentally clarified details of the electronic state.

Ag is a noble metal with an electron configuration of [Kr]4$d^{10}5s^1$. Individual neutral atoms have one unpaired electron, thereby exhibiting paramagnetism, which is due to a spin-orbit coupling, whereas bulk metallic silver is diamagnetic. Furthermore, the reduced silver clusters with the nano-size are often paramagnetic \cite{Her80,Gro85,Mic86,Lee89} and appear to provide a bridge between the limits of the isolated atoms and bulk. With a few exceptions, spin paramagnetism is assumed due to the number of unpaired electrons which can lead to the electronic size effects.

An inherent advantage, particularly in small metal clusters, is due to the fact that the unpaired electrons are assumed to occupy molecular orbitals delocalized throughout the overall cluster atoms, can be expressed surprisingly well by a simple {\it tight-binding approximation} leading to a simple Fermi level which is being used extensively for electronic states confined in a low dimension \cite{Mar79}. A few electronic levels near the highest occupied electronic state of energy are of great interest, and their interactions/repulsions with the lowest unoccupied electronic state result in different energy level distributions depending upon the symmetry of the electron Hamiltonian. The energy difference between HOMO and LUMO is the energy gap $\delta$ (or Kubo gap) which will be comparable to other energies such as the thermal energy $k_BT$, the electron Zeeman $\mu_{\rm B}H$ and the nuclear Zeeman $\mu_{\rm N}H$ at low temperatures. The Fermi level lies in the energy gap, while the electron excitation through the energy gap can be extensively studied by applying the external magnetic field \cite{Hal86}. Time-dependent DFT calculations revealed that a transfer/excitation of $5s$-electron from HOMO to LUMO is mainly responsible for the absorption spectra of very small Ag clusters \cite{Zha06}.

The surface effects arises from the boundary condition for the electronic wavefunction at the surface. The relative merit of the surface effects is the fraction of atoms at the cluster surface, and a low mean coordination number basically implies a large fraction of surface atoms. The surface atoms make a crucial contribution to the cluster total energy which leads to a surface reconstruction due to a strong tendency of surface energy minimization. Such structural changes affect strongly the electronic state or wavefunction. The structural characteristics of metal nano-clusters are more important to structure-sensitive properties such as magnetism \cite{Agu05}.

Theoretical calculations on the evolution of magnetism of bare Ag clusters as a function of size and on the effects of cluster symmetry have only recently become available \cite{Per07}. The neutral, free Ag$_6$ cluster was predicted to be diamagnetic and of planar triangular $D_{3 \rm h}$ symmetry. Furthermore, an Ag$_{13}$ cluster of icosahedral symmetry was predicted to have the highest magnetic moment per Ag atom. This work generally indicated that the electronic structure, the coordination number, and the specific symmetry of small Ag clusters are fundamental factors in the development of magnetism. The only experimental work revealed the negative magnetic susceptibility due to diamagnetism of the auto-reduced Ag/NaA, and the value was on the order of $-2.5\times10^{-7}$ to $-4.5\times10^{-7}$ emu/mol. Initially, the formation of paramagnetic Ag$^{2+}_3$ clusters was proposed, but a trace of paramagnetism was only due to impurity of Fe$^{3+}$ in the support lattice \cite{Tex86}.

The preparation of paramagnetic metal clusters in the support pores is often fulfilled by utilizing the chemical methods of ion-exchange, oxidation and reduction. This strategy can produce a large variety of structures, i.e. paramagnetic atoms, ions, and two- or three-dimensional frameworks, and some of which can exhibit a high spin state and some others have a low spin state or diamagnetic behavior. Size-selected Ag clusters are objects of great scientific interest on account of their unexplored physical, electronic and chemical properties in particular. The synthesis of silver clusters at a loading of 12\% (wt.) in NaA zeolite leads to a well-resolved EPR spectrum which is compatible with a structure consisting of six equivalent Ag atoms \cite{Amg09,Bal09}. Depending on the metal loading, Ag$^0$, Ag$^{\rm n+}_3$ and Ag$^{\rm n+}_4$ clusters are observed by EPR as well \cite{Amb09,Amg08}, but in all cases calibration of the signal reveals that it represents $<1\%$ of all the exchanged silver in the sample, the majority species is EPR-silent. It is of interest to know whether the silent fraction involves diamagnetic or high-spin states. However, the present work is restricted to monitoring the static magnetization of the nominal Ag$^+_6$ cluster. Magnetization measurements reveal the overall magnetic properties of the whole silver cluster-support system, whereas EPR selectively probes the paramagnetic Ag$^+_6$ clusters in the field domain.

XAS is a very useful and powerful tool for providing full information content of the local structure of the photon absorbing atom. The local structure is determined by analyzing the oscillatory $\chi(k)$ function of the absorption fine structure spectra. An average coordination number of Ag atoms in the nearest neighboring shells with different distances is a main structural parameter reflecting the mean size of the supported Ag clusters. The coordination in the first two shells is sufficient to estimate the cluster size, but might be not enough to determine the cluster shape \cite{Jen99}.

Thermodynamically, clusters spontaneously aggregate to larger particles ($d > 10$ nm), thereby minimizing surface energy. It is an advantage that the support pores/cages restrict the interactions of the clusters and provide a practical means of preventing the cluster aggregation. However, it may be difficult to separate the size-dependent effects of the supported Ag clusters from the inevitable effects of the cluster-support interactions.

It is ultimately emphasized that the surface effects and electronic size effects are inextricably correlated, but are not related by a simple scaling law. This means that these size-dependent effects are not perfectly scalable or smooth with each other, since they can depend not only on the shape but also on the delocalized molecular orbital structure of the clusters at the same time.

\section{Experimental}

The NaA (Si/Al = 1) zeolite was supplied by CU Chemie Uetikon AG in Switzerland. Zeolite samples were heated up in air at a rate of 0.5 K min$^{-1}$ to 773 K where they were kept for 14 hours in order to burn off any organic impurities. Subsequently, 7 g of the heated sample was washed by stirring in 150 ml bi-distilled water containing 40 ml NaCl (10\%) solution and 2.76 g of Na$_2$S$_2$O$_3\cdot$5H$_2$O salt at 343 K. This washing processes was repeated at least nine times. The washed sample was dried in air at 353 K for 24 hours.

Ag/NaA samples were prepared in a flask containing 2.25 g of pre-treated zeolite by aqueous ion-exchange with 50 ml 50 mM AgNO$_3$ solution (ChemPur GmbH in Germany, 99.998\%) by stirring at 343 K in the dark for 24 hours. The ion-exchanged sample was filtered and rinsed with deionized water several times, and dried in air at 353 K overnight. Chemical analysis by atomic absorption spectroscopy (AAS) demonstrated that the ion-exchange reaction leads to a silver loading of ca. 12\% (wt.).

Oxidation was performed under a gas stream of O$_2$ (Westfalen AG in Germany, 99.999\%) with a flow rate of 17 ml min$^{-1}$ g$^{-1}$ from room temperature up to 673 K using a heating rate of 1.25 K min$^{-1}$ where it was kept for an additional hour. While the sample was held at the final temperature, the residual O$_2$ gas in the reactor was purged by N$_2$ (Westfalen AG, 99.999\%) gas for 1 hour. Subsequently, the sample was sealed and kept at 673 K overnight.

After cooling the sample, reduction was performed in a flow of H$_2$ gas (Westfalen AG, 99.999\%, 16 ml min$^{-1}$ g$^{-1}$) at room temperature or below for 20 minutes, which leads to the stabilization of Ag$^+_6$ clusters. Alternatively, D$_2$ (Westfalen AG, 99.0\%) reduction was carried out under a static gas pressure of 500 mbar. The paramagnetic Ag cluster containing samples were also subjected to a second reduction under static conditions of 500 mbar gas pressure. This is because XAS experiments prompted us to assume that this second treatment leads to a more complete reduction of the Ag cluster species while the first treatment just leads to partial reduction.

Continuous wave EPR spectra were recorded on a Bruker EMX X-band spectrometer operating at a microwave frequency of 9.5 GHz, using a modulation amplitude of 8 G and a microwave power of 1 mW. Analysis was by spectrum simulation with EasySpin based on a MATLAB numerical function \cite{Sto06}.

Magnetic property measurements were performed using a Superconducting Quantum Interference Device (SQUID) magnetometer (Quantum Design MPMS-7, XL7) as a function of an applied external magnetic field with a maximum strength of 7 T (Tesla) in the temperature range of 1.8 K to 300 K. The reduced samples were wrapped in Teflon tape and pressed into tight pellets under nitrogen or argon gas in a glove box. Magnetic susceptibility measurements of the wrapped samples were performed at 2000 Oe (0.2 T) of a fixed magnetic field as a function of temperature. Alternatively, the reduced sample were transferred into NMR quartz tubes (outer diameter about 5 mm) under nitrogen or argon gas in a glove box, and the tubes were sealed with stopcocks for vacuum treatments. Magnetic susceptibilities were measured for these samples at 1000 Oe, 3000 Oe, and 10000 Oe of a fixed magnetic field. All the experimental data were corrected by subtracting the magnetic susceptibility of an identical amount of silver-free NaA zeolite to obtain the susceptibility of the nominal Ag$^+_6$ cluster alone.

XAS experiments at the Ag $K$-edge (25.514 keV) were performed at the Swiss-Norwegian Beam Line BM01 of the Storage Ring at the ESRF. A Si (311) single crystal was used as channel-cut lens for obtaining a monochromatic X-ray beam. Energy and current (intensity) of the storage ring were 6.0 GeV and 200 mA during the operation. X-ray absorption spectra were recorded in transmission mode at room temperature using ionization chamber detectors. Energy calibration was performed with a silver metal foil. The isolation of the $\chi(k)$ function from experimental EXAFS data was performed using the XDAP software package \cite{Var95}. Fourier transform of the $\chi(k)$ function leads to the determination of the radial distribution function defined in $R$-space in terms of the distance from the absorber atom. All details about data extraction and analysis are described elsewhere \cite{Kon00}.

\section{Theoretical background}
The magnetic susceptibility of diamagnetic solids is negative. The dimensionless scalar susceptibility is related to the magnetic permeability via
\begin{equation}
\label{Vol_Sus}
\mu = (1 + \chi)\mu_0
\end{equation}
where $\chi$ is the dimensionless {\it volume susceptibility} or {\it bulk susceptibility}. Experimental data are often quoted as {\it molar susceptibility}, $\chi_{\rm mol}$ in units of emu/mol (or cm$^3$/mol, or m$^3$/mol). The paramagnetic contribution to $\chi_{\rm mol}$ is frequently expressed using the Curie-Weiss formula
\begin{equation}
\label{Sus}
\chi_{\rm mol} = \frac{M}{H} = \chi_0 + \frac{C_{\rm m}}{T - \theta}
\end{equation}
where $M$ is the magnetization in electromagnetic units (emu/mol), $H$ is the applied field in Oersted (Oe), $C_{\rm m}$ is the Curie constant in emu$\cdot$K/mol, $\theta$ is the Curie-Weiss temperature in Kelvin (K). All units are in CGS.

This susceptibility of paramagnetic matters can also vary inversely with temperatures
\begin{equation}
\label{Inv_Sus}
\chi^{-1} = \ [\chi_{\rm mol} - \chi_0]^{-1} = \frac{T - \theta}{C_{\rm m}}
\end{equation}
where $\chi_0$ is a temperature independent contribution, normally the diamagnetic ($\chi < 0$) or for many metals a Pauli-paramagnetic ($\chi > 0$) susceptibility.

The magnetic susceptibility of a system is usually converted into the effective magnetic moment, $\mu_{\rm eff}$, derived from $C_{\rm m}$ using
\begin{equation}
\label{Eff_Mom}
C_{\rm m} = \frac{N_{\rm A}\mu_{\rm eff}^2}{3k_B} = \frac{N_{\rm A}g^2J(J+1)\mu_{\rm B}^2}{3k_B}
\end{equation}
where $N_{\rm A}$ is Avogadro's number, $k_B$ is Boltzmann's constant, $J$ is the total (spin plus orbital) angular momentum, and $g$ is a spectroscopic splitting factor.

The field dependence of magnetization of a classical paramagnet is described by the classical Langevin function \cite{Blu01},
\begin{equation}
\label{Mag}
M(H) = N \mu \Big [\coth \Big (\frac{\mu H}{k_B T} \Big ) - \frac{k_B T}{\mu H} \Big ]
\end{equation}
where $N$ is the total number of atoms per unit volume, $T$ is the temperature, $H$ is the magnetic field, $\mu$ is the magnetic moment per atom, and $k_B$ is Boltzmann's constant. However, this is a semi-classical treatment of paramagnetism and corresponds to a $J = \infty$ system. It is therefore important that the classical moment is replaced by a quantum spin $J$ in quantum mechanical systems. Subsequently, a specific function can be derived and described via
\begin{equation}
\label{Sat_Mag}
\frac{M}{M_{\rm S}} = \frac{\langle m_J \rangle}{J} = \tanh y
\end{equation}
where $y = \mu_{\rm B} H/k_B T = g J \mu_{\rm B} H /k_B T$ for a $J$ = 1/2 and $g$ = 2 system, and $M_{\rm S}$ is a saturation magnetization.

Atomic Ag has an electron configuration of [Kr]4$d^{10}5s^1$ and a spin doublet ground state. In bulk Ag, the 5$s$ electrons outside the closed $d$-shell can be considered practically free and are responsible for electrical conduction. These $s$ electrons exhibit a paramagnetic contribution to susceptibility, and the total susceptibility contains a diamagnetic contribution from the $d$ electrons and from lower shells. At room temperature the molar susceptibility is $-19.5 \times10^{-6}$ emu/mol for bulk metallic Ag \cite{Per07}.

In EXAFS measurements the oscillatory $\chi(k)$ function is a summation over all coordination shells around an absorbing atom and described via
\begin{eqnarray}
\nonumber
\chi (k) &=& \sum_{j} \frac {S_0^{2} N_j}{k R_j^2} f_j(k, R_j) \exp (-2 \sigma _j^2 k^2) \\
&&
\exp \Big ( \frac {-2 R_j}{\lambda (k)} \Big ) \sin [2kR_j + \varphi _j(k)]
\label{EXAFS_func}
\end{eqnarray}
where $R$, $N$ and $\sigma$ are the inter-atomic distance, coordination number and root mean square displacement (Debye-Waller factor), respectively, for each atomic pair. The back-scattering amplitude $f_j(k)$ and the phase factor $\varphi_j(k)$ of the sine function are both functions of $k$. Equation (\ref{EXAFS_func}) is called the EXAFS function which provides information about the local structure about the absorbing atom \cite{Kon00}.

\section{Results and discussion}

\subsection{Continuous wave X-band EPR measurements}

\begin{figure}[ht]
\centering
\includegraphics[width=0.675\columnwidth]{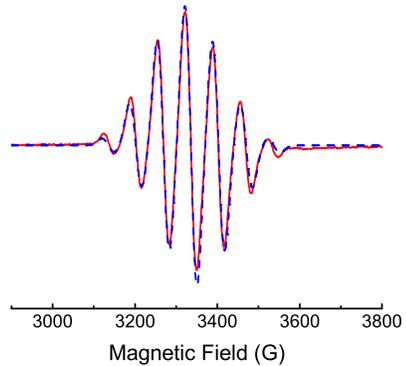}
\caption{X-band EPR spectrum of the Ag$_6^+$ cluster recorded at 20 K (solid line) after H$_2$ reduction at 278 K for 20 minutes, and corresponding simulation (dashed line) based on isotropic hyperfine parameters for 6 equivalent Ag nuclei. The simulation takes into account the statistical distribution of Ag isotopes.}
\label{fig:Cluster}
\end{figure}

The X-band EPR spectrum of the hydrogen reduced 12\% (wt.) Ag/NaA is illustrated in Figure \ref{fig:Cluster}. This well-defined isotropic signal with $g_{\rm iso}$ = 2.028 is assigned to the reduced Ag$^+_6$ cluster. $g_{\rm iso}$ is significantly higher than the free electron $g_{\rm e}$ value, which normally indicates admixture of a transition metal $d$-orbital into spin density distribution \cite{Dyr97}. In a simple picture, a hole in the atomic 4$d$-shells would imply a cluster charge that exceeds +6. More likely is a significant admixture of $d$ atomic orbitals in the molecular cluster orbital containing the unpaired electron, or an angular momentum of the half-filled molecular orbital that has a similar effect as the atomic $d$-orbitals.

The unpaired spin density is delocalized equally over all Ag atoms of the cluster, which leads to an identical and isotropic hyperfine coupling with all nuclei. It implies that there are allowed transitions of $\Delta m_S = \pm1$ between $m_S$ = - 1/2 and $m_S$ = + 1/2 states of electron spin angular momentum. Since the bulk Ag has the electronic configuration of Kr[4$d^{10}5s^1$], the paramagnetic property of small Ag clusters derives from the odd-number of 5$s$ valence electrons in the open-shell configuration.

The observed multiplet must be explained as a superposition of spectra of clusters with statistical distribution of Ag isotopes. The statistical isotope distribution determines mainly the linewidth. The isotopes $^{107}$Ag and $^{109}$Ag have $I$ = 1/2 spins and possess almost the identical abundance (51.839\% and 48.161\%). $^{109}$Ag has a nuclear magnetic moment that is 15\% larger than that of $^{107}$Ag, and at the given spectral resolution the difference in magnetic moment is negligible. In the experiment the difference is not resolved, but contributes to the linewidth. The experimental $a_{\rm iso}$ = 67.0 G is to be compared with an average atomic value of 702 G of the $^{107}$Ag and $^{109}$Ag isotopes \cite{Mor78}. Thus, the 5$s$ orbital contribution is about 10\% per nucleus or 60\% for all Ag nuclei together. It is well known from the free-electron approximation that if size of clusters is small, there is a possibility of a partial fraction of the $s$ orbital radial $|\psi_0(0)|^2$ wavefunction to remain outside of a quantum sphere or wall \cite{Mar79}, leading to a significant missing fraction of the total spin density. The characteristic hyperfine structure remains well-resolved at 200 K and is still observable at room temperature. It should be noted that very little $g$ or hyperfine anisotropy is present. This suggests that the cluster has close to spherical symmetry, and it essentially excludes that the spectrum is due to alternative structures with six equivalent atoms, e.g. a planar hexagon. it is supposed to preserve the equivalence of the Ag nuclei without distortions in the cluster structure. The numerical spectrum simulation was performed applying the experimental spin Hamiltonian parameters for six equivalent nuclei.

The double integral of the EPR signal is directly proportional to the number of unpaired spins in a sample. Calibration with a standard sample (ultramarine blue was used in this case) and assuming $S=1/2$ yields a value of 4.8$\times$10$^{16}$ spin g$^{-1}$, which accounts for only 0.044\% of all silver atoms in the sample. This means that most of the silver is EPR silent, i.e. diamagnetic or unobservable high spin states. The inverse signal amplitude, plotted in a temperature range of 4.2 K to 140 K, exhibits convincing Curie behavior (not shown), in agreement with non-interacting electron spins. A count of paramagnetic spin is consistent with the fact that only one of 24 $\beta$-cages in Ag/NaA is occupied by a nominal Ag$^+_6$ cluster. It was reported that paramagnetic Ag$^+_6\cdot$8Ag$^+$ cluster was formed in the $\beta$-cage of Ag$_{12}$/NaA by $\gamma$-irradiation at 77 K \cite{Mor87}. Irradiation converts out only $\approx$ 1 of 5000 EPR silent clusters into the paramagnetic Ag$^+_6\cdot$8Ag$^+$ cluster, even though the exchanged Ag$^+$ replace completely all 12 Na$^+$ in the unit cell of NaA (Ag$_{12}$/NaA prepared from 1 M AgNO$_3$ solutions). 

\subsection{EXAFS measurements}

X-ray absorption spectroscopy (XAS) offers an excellent potential for estimating the mean size and local structure of atoms in the small metal clusters \cite{Oud00,Vaa96}. The highest fraction of the surface atoms implies a low mean coordination number which is accompanied by a direct consequence of the bond length contraction. This is taken as experimental evidence for the existence of the nano-size effects.

\begin{figure}[ht]
\centering
\includegraphics[width=1\columnwidth]{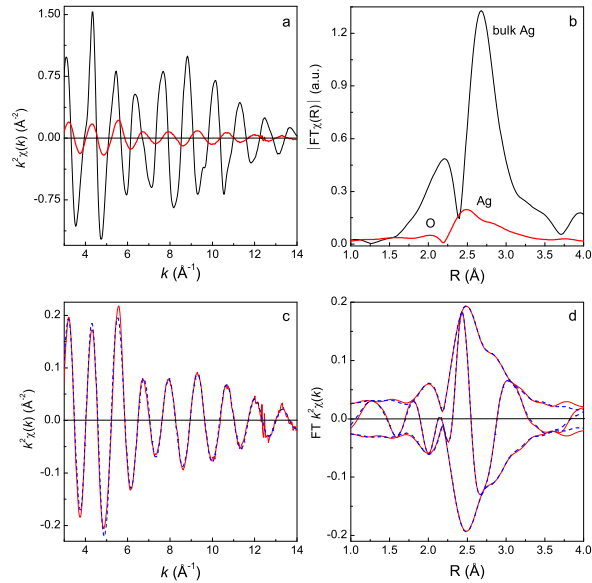}
\caption{EXAFS spectra of the H$_2$ doubly reduced 12\% (wt.) Ag/NaA: a) Comparison of the $k^2$-weighted $\chi (k)$ function ($k^2$, $\Delta k=3.0-14$ {\AA}$^{-1}$) of the experimental EXAFS data for Ag clusters (red line) with that of bulk Ag (black line). b) Comparison of the Fourier transform of the $k^2$-weighted $\chi (R)$ ($1.0< R<4$ {\AA}) radial function of experimental data (red line) with that of bulk Ag (black line). c) $k^2$-weighted $\chi (k)$ function ($k^2$, $\Delta k=3.0-14$ {\AA}$^{-1}$) of the experimental data (red line) and best fit (blue dashed-line) including the first and second Ag-Ag shell, and a single Ag-O shell. d) Real and imaginary part of Fourier transform of the $k^2$-weighted $\chi (k)$ function ($k^2$, $\Delta k=3.0-14$ {\AA}$^{-1}$) of the experimental data (red line) and best fit (blue dashed-line) including the first and second Ag-Ag shell, and a single Ag-O shell.}
\label{fig:XAS}
\end{figure}

Absorption fine structure spectra for a doubly reduced supported silver sample are illustrated in Figures \ref{fig:XAS}. The oscillatory $\chi(k)$ function is completely different from that of bulk Ag in that it is of much lower amplitude, has a longer wavelength and decreases more rapidly on increasing $k$ values. The significant loss in intensity of $\chi(k)$ is due to the extremely small size of the Ag clusters (see Figures \ref{fig:XAS}(a) and (c)). Fourier transforms of the isolated $\chi(k)$ function were performed using the $k^2$-weighting in the $k$ range from 3 to 14 {\AA}$^{-1}$. The isolated shell contributions are represented as Ag-Ag and Ag-O peaks in the Fourier transform ($R$-space) at 2.5 and 2.1 {\AA} in Figure \ref{fig:XAS}(b) and (d), respectively. Note that no phase or amplitude corrections were performed during the Fourier transform. A heavy scatterer such as Ag experiences a phase-shift by photo-electron scattering that shifts the peak in Fourier space to a lower value of $R$. This is clearly seen in Figure \ref{fig:XAS}(b) and its Fourier transform.

The absorbing Ag atom of the doubly-reduced cluster has an average coordination of $N \approx 3.35\pm0.05$ at an average distance of 2.76 {\AA} in the second Ag-Ag shell, while the first Ag-Ag shell at 2.69 {\AA} has $N \approx 0.55\pm0.05$ on average. When a second shell of atoms is introduced, the coordination number $N$ of the first shell decreases \cite{Vaa96,Nay97}. This represents a surface reconstruction due to energy minimization of the clusters. It is emphasized that the use of the two-shell model for Ag-Ag contribution improves the fit results significantly. The total Ag-Ag coordination number is $4.00\pm0.07$. All atoms belong to the cluster surface. This makes immediately clear that silver is present quantitatively as small clusters, as any contribution of nanoparticles would increase $N$ much closer to its bulk value of 12. A regular 6-atom octahedron should give an overall $N$ of 4.6, since the fifth backscattering atom sits across the body diagonal, ca. 4.0 {\AA} from the absorbing atom and is thus not contributing to the scattering amplitude in the discussed range. Neutral 6-atomic silver clusters are predicted to be planar triangles of $D_{\rm 3h}$ symmetry \cite{Per07}, which leads to an average $N$ of 3.0. It should be noted that in the same work the most stable structure of neutral Ag$_7$ is predicted to be a pentagonal bipyramide, which has an average coordination number of 4.3, essentially identical with the above experimental value of 4.20.

The obtained experimental Ag-Ag distance of 2.76 {\AA} on average reveals a contraction of the bond lengths in comparison to bulk Ag (2.889 {\AA}), which is a typical nano-size effect. A separation between HOMO and LUMO increases with decreasing the bond length of small Ag clusters \cite{Zha06}. Furthermore, an oxygen shell is detected at 2.27 {\AA} with a coordination of $N \approx 0.70\pm0.05$. These may be from the zeolite framework to which the cluster is loosely attached.

\begin{figure}[ht]
\centering
\includegraphics[width=0.675\columnwidth]{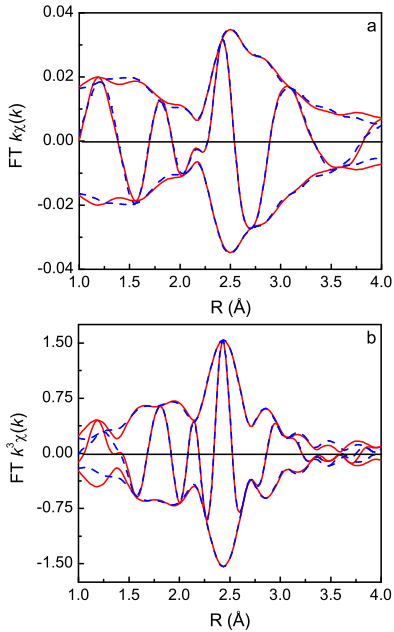}
\caption{EXAFS spectra of the H$_2$ reduced Ag/NaA: Fourier transform of a) $k^1$-weighted and b) $k^3$-weighted $\chi (k)$ function ($k^1$ and $k^3$, $\Delta k$ = 3.0 - 14 {\AA}$^{-1}$) of experimental EXAFS data (red line) with the corresponding fit (blue dashed-line) including the first and second Ag-Ag shells, and single Ag-O shell.}
\label{fig:XAS_1}
\end{figure}

The Debye-Waller factor ($\Delta \sigma^2$) and the coordination number ($N$) are highly correlated in a fit. In order to check a proper set of parameters, the fits to the real and imaginary parts of the $k^1$- and $k^3$-weighted Fourier transforms are compared, with the contributions consisting of the Ag-Ag and the Ag-O shells isolated in the range 1 $ < R < 4$ {\AA} (see Figure \ref{fig:XAS_1}(a) and (b)). The $k^3$-weighted data enhance the sensitivity for the heavy scatterers. The real part of the Fourier transform is dependent on $N$ and disorder, while the imaginary part is extremely sensitive to the inter-atomic Ag-Ag distance and therefore used to judge the quality of a fit \cite{Kon00}. The $k^3$-weighted Fourier transform of the oscillatory $\chi(k)$ function of the reduced Ag cluster in Ag/NaA is almost symmetric, which is a strong indication of larger contributions by overlapping Ag-Ag shells. The appearance of multiple satellite peaks in the Fourier transform would describe a single absorbing backscattering pair. This is therefore excluded for the present Ag clusters.

For comparison with the doubly reduced sample Figure \ref{fig:XAS_1}(b) shows the best fit of the Fourier transform of data after a single reduction process. The silver has an average Ag-Ag coordination of $N \approx 3.50\pm0.05$ at an average distance of 2.77 {\AA} in the second shell, and $N \approx 0.40\pm0.05$ at a distance of 2.69 {\AA} in the first shell. The coordination of oxygen is $N \approx 1.60\pm0.05$ at 2.22 {\AA}, more than double of what it is after an additional reduction step. It thus seems that the cluster is indeed incompletely reduced. The agglomeration of the cluster is either incomplete, or less spherical, or the oxygen is partly interspersed between the silver atoms. The theoretical fit matches nicely the imaginary and real parts of the Fourier transform, though a small asymmetry of the spectrum is induced by the presence of a light back-scatterer below 1.6 {\AA} in $R$-space in Figure \ref{fig:XAS_1}(a), perhaps O atoms of the support. Furthermore, the root mean-square displacement ($\Delta \sigma^2 = 0.012$) is relatively large and may be taken as evidence of structural disorder or temperature activated fluctuations because all measurements were carried out at room temperature. Upon evacuating the samples, there were no significant changes in the structural parameters of the reduced silver clusters.

\subsection{SQUID measurements}

\subsubsection{Magnetization}

The experiments on the field dependence of magnetization $M(H)$ were performed up to 70 kOe in steps of 2500 Oe at different selected temperatures. Magnetization of an equivalent amount of the silver-free zeolite (or before ion exchange) but with the same treatment was subtracted.

At a low temperature regime of 1.8 K, $M(H)$ increases, but exhibits a nonlinear dependence on the magnetic field strength. Assuming an $S=1/2$ state, the $M(H)$ curve was fitted using the modified and extended version of Equation (\ref{Sat_Mag}), i.e. $M(H) = N \mu_0 \mu_{\rm B} \tanh(\mu_0 \mu_{\rm B}H/k_BT) + \chi_0H$. The first term accounts for a nonlinearity of $M(H)$ due to alignment of the unpaired spin by a magnetic field at a low temperature. This term includes the effect of the magnetic moment on the Curie-Weiss dependence regarding the odd-electron susceptibility \cite{Hal86}. The second term describes a significant linear contribution of a temperature-independent susceptibility $\chi_0$ to the $M(H)$ curve. In some other cases with a few exceptions, this also describes an appreciable linear contribution to $M(H)$ determined by a field-independent susceptibility $\chi_0$ \cite{Vav06}. It is emphasized that the subtraction/correction of a constant and temperature independent component (e.g. $\chi_0$) for magnetization excludes the determination of any {\it diamagnetic} or {\it Pauli paramagnetic contribution} \cite{Cor07}.

\begin{figure}[ht]
\centering
\includegraphics[width=0.725\columnwidth]{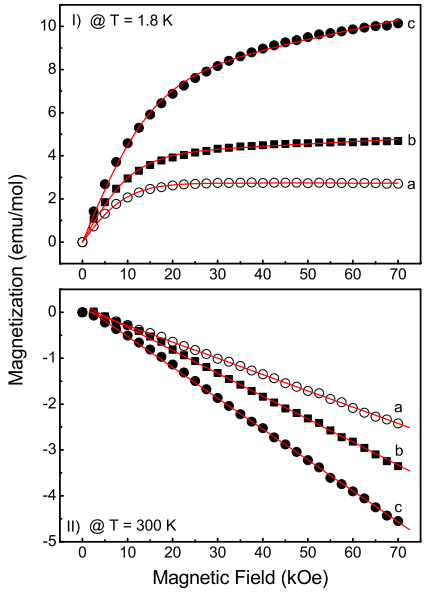}
\caption{Magnetic field-dependence of molar magnetization (in emu/mol) for reduced clusters in 12\% (wt.) Ag/NaA samples measured in a range of 0 to 70 kOe: I) at at 1.8 K  and II) at 300 K; (a) H$_2$ reduced (b) H$_2$ desorbed (c) D$_2$ reduced samples. The red lines are the theoretical fits to the field-dependence using the extended version of Equation (\ref{Sat_Mag}) at 1.8 K (I) and of the linear behavior describing a temperature-independent $\chi_0$ contribution at 300 K (II).}
\label{fig:sat}
\end{figure}

Three magnetization curves recorded at 1.8 K are displayed in Figure \ref{fig:sat}, and the measurements contain a more reliable magnetic information of the cluster ground state. The curvature does not depend on the total magnetization, therefore, two independent pieces of information can be obtained, the magnetic moment ($\mu$) per cluster and the fraction ($f$) of magnetic atoms per mole in the sample. Since the number of atoms is 6 in each nominal Ag$^+_6$ cluster, N can be taken as the number of clusters in a mole and the number of Ag atoms constituting the cluster can be also expressed in a mole ($f={\rm N}^{\rm cl}_{\rm Ag}/N_{\rm A}=6{\rm N}/N_{\rm A}$). All fit parameters and calculated magnetic moments are summarized in Table \ref{tab:Magnetization}.

The magnetic moment per nominal Ag$^+_6$ cluster, is 3.0 $\mu_{\rm B}$, 4.0 $\mu_{\rm B}$, and 4.8 $\mu_{\rm B}$ on average, respectively. Only 1.52\%, 0.6\%, and 0.4\% of the exchanged silver atoms contribute forming a part of magnetically active Ag$^+_6$ clusters in the deuterium-reduced, hydrogen-desorbed and hydrogen-reduced samples. Furthermore, the values of the magnetic moment probably imply that there is an odd electron effect of the cluster because the difference is almost 1 $\mu_{\rm B}$ per cluster \cite{Yam03}. 

At higher temperatures magnetization is not consistent with any Curie type behavior. This is probably not that surprising, because Curie's law is only valid in the high-temperature/low-field regime. Since the $M(H)$ curves are linear but negative at 300 K (see Figure \ref{fig:sat}(II)), the slope of the linear fit gives an access to the value of $\chi_0$ at any magnetic field strength. It is however emphasized that this negative linear relation of $\chi_0$ can include all the temperature independent contributions such as van Vleck and diamagnetism of the ion cores, as well as other spurious contributions of the sample holder and errors in the substrate subtraction \cite{Sal07}. Thus, the fit derived $\chi_0$ cannot represent a pure diamagnetic contribution to magnetism, and the sign of the values has no physical meaning.

\begin{table}[ht]
\small
\caption{Magnetic moments and fit parameters derived from experimental $M(H)$ data for different reduction conditions. Measurements as a function of applied magnetic field were performed at a fixed temperature of 1.8 K.}
\label{tab:Magnetization}
\begin{tabular*}{0.475\textwidth}{@{\extracolsep{\fill}}lccc}
\hline
Parameters 	                                          & D$_2$ red      & H$_2$ des      & H$_2$ red        \\
\hline 
$\mu_{\rm cl}=6\mu_{\rm at}$ ($\mu_{\rm B}$/Ag$^+_6$) & 3.018          & 4.122          & 4.884            \\ 

$\mu_{\rm at}$ ($\mu_{\rm B}$/Ag atom)                & 0.503          & 0.687          & 0.814            \\ 

N (clusters in a mol)                                 & 14.141         & 5.622          & 3.283            \\ 

$M_{\rm S}$ = N$\mu_{\rm at}$ (emu/mol)               & 7.113          & 3.862          & 2.672            \\ 

$f$ = 6N/$0.927\times 10^{-20}N_{\rm A}$              & 0.015          & 0.006          & 0.004           \\ 

$f \mu_{\rm at}$ ($\mu_{\rm B}$/Ag atom)              & 0.008          & 0.004          & 0.003           \\ 

$\chi_0(\times10^{-6})$ (emu/mol)                     & -40.0          & -10.0          & -0.064           \\ 

$\chi_0(\times10^{-6})$ (emu/mol)                     & -67.31         & -49.95         & -35.4(5)         \\
\hline 
\end{tabular*}
\end{table}

The surface is highly important to metal nano-clusters, the surface potential has an effect on the electronic wavefunction. A high abundance of clusters with a certain size (so-called magic numbers) is often related to a particular stabilization in closed-shell geometric or electronic structures. The only magic cluster with six equivalent atoms is a regular octahedron. However, paramagnetic clusters do not have a closed shell, thus often adopting a lower symmetry. In fact, all Ag atoms belong to the surface of the nominal Ag$^+_6$ cluster, while no atoms are in the core.

By theoretical predictions all Ag atoms of bare free Ag$_3$ to Ag$_{12}$ clusters are at the surface, and even-numbered clusters are diamagnetic because of the even number of electrons in the closed-shell electronic configuration \cite{Per07}. Furthermore, theoretical calculations predicted a sextet high spin ground state for free icosahedral neutral Ag$_{13}$, a doublet ground state for all the smaller odd-atomic neutral clusters, and singlet ground states for even-atomic clusters with less than 14 atoms \cite{Per07}. The larger clusters were predicted to be "magnetically fluxional", with low energy excited magnetic states. For positively charged clusters such calculations have not yet been performed, and in the real system the interaction with the zeolite lattice will play a role as well, which opens the possibility for a wide variety of clusters which differ in their detailed properties even though they have a narrow size distribution. A coordination of 4 atoms in the second shell and of one in the first shell is strongly reduced compared with the bulk metal, suggesting that a degeneracy is lifted by small irregularity in an octahedral symmetry of the reduced Ag$^+_6$ cluster.

\subsubsection{Magnetic susceptibility}

The temperature dependence of the dc susceptibility was measured at a magnetic field strength of 2000 Oe after a number of different reduction treatments of silver clusters in 12\% (wt.) Ag/NaA.

\begin{figure}[ht]
\centering
\includegraphics[width=0.85\columnwidth]{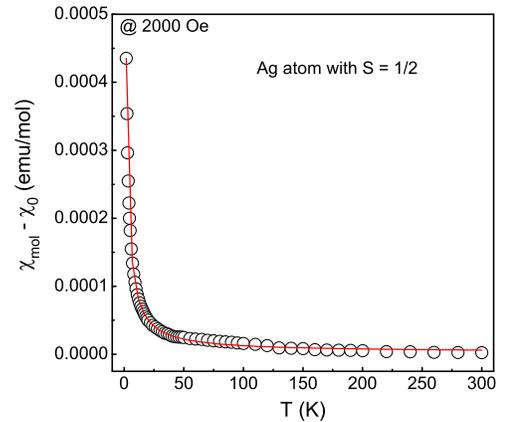}
\caption{Molar magnetic susceptibility $\chi(T)$ as a function of temperature for hydrogen-desorbed cluster in 12\% (wt.) Ag/NaA at a magnetic field of 2000 Oe. The red line representing the fit to the temperature dependence is calculated using Equation (\ref{Sus}), and the fit values of $\chi_0$ are subtracted to obtain the contribution of $\chi(T)$ to paramagnetic susceptibility.}
\label{fig:sus}
\end{figure}

\begin{figure}[ht]
\centering
\includegraphics[width=0.75\columnwidth]{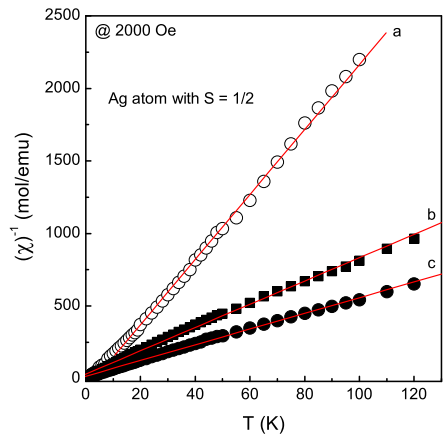}
\caption{Inverse magnetic susceptibility (magnetization divided by the field in Oe) in mol/emu per Ag atoms in Ag/NaA as a function of temperature in the range of 1.8 K to 120 K) at a field of 2000 Oe: (a) H$_2$ reduced (b) H$_2$ desorbed (c) D$_2$ reduced. The slope of the curves is proportional to 1/$C_{\rm m}$, which is used for calculating the effective magnetic moment. The inverse susceptibility $(\chi)^{-1}$ is inverse of $\chi(T)$ plotted in Figure \ref{fig:sus}.}
\label{fig:inv}
\end{figure}

The temperature dependence of $\chi(T)$ can be fitted using Equation (\ref{Sus}), which contains a sum of a Curie-Weiss term and a temperature-independent term $\chi_0$, and the values of the three-parameter fit are given in Table \ref{tab:Mag_val}. By subtracting a temperature-independent $\chi_0$ contribution, the static susceptibility $\chi(T)$ of the hydrogen-desorbed cluster demonstrates a nearly ideal paramagnetic Curie-Weiss dependence with $|\theta|$ = 0.5 K. The fit result is displayed in Figure \ref{fig:sus}. Furthermore, the inverse susceptibility 1/$\chi$ is plotted in a temperature range of 1.8 K to 120 K in Figure \ref{fig:inv}. The Curie-Weiss temperature $\theta$ is very close to zero, ca. $|\theta|<1$ K, as demonstrated by the straight line. This supports the implicit assumption that there is no interaction between the magnetic moments of neighboring clusters. 

Using Equation (\ref{Eff_Mom}) where $N_{\rm A}$ is Avogadro's number ($6.02\times 10^{23}$) and $k_B$ is Boltzmann's constant ($1.38\times 10^{-16}$ erg/K), the molar Curie constant $C_{\rm m}$ derived from the $\chi(T)$ fit was applied to calculate the effective magnetic moment $\mu_{\rm eff}$ per Ag atom which will be in units of erg/Oe. The effective moment is usually reported in units of Bohr magnetons (denoted $\mu_{\rm B}$). Dividing by $0.927\times 10^{-20}$ (erg/Oe)/$\mu_{\rm B}$ will give $\mu_{\rm eff}$ in units of $\mu_{\rm B}$ (see Table \ref{tab:Mag_val}). However, the actual formula is only valid for an ideal system of {\it identical clusters}, each comprise 6 atoms with moment $\mu_{\rm B}$. All Ag clusters cannot be identical as for the mean size and magnetic activity, only a fraction $f$ of Ag atoms forms a part of magnetically active clusters which bear an effective or permanent magnetic moment. By utilizing $f\mu^2_{\rm cl}/6=\mu^2_{\rm eff}$, the magnetic moment $\mu_{\rm cl}$ per Ag cluster can be calculated applying the effective moment $\mu_{\rm eff}$ derived from the $\chi(T)$ fit and $f$ deduced from the $M(H)$fit. This yields 2.6 $\mu_{\rm B}$, 2.9 $\mu_{\rm B}$, and 3.1 $\mu_{\rm B}$ per Ag cluster, respectively (see Table \ref{tab:Mag_val}). It is therefore considered that the $\chi(T)$ fit provides nearly consistent data with the $M(H)$ fit given in Table \ref{tab:Magnetization}. The strength of the applied magnetic field is very decisive to induce the susceptibility \cite{Hat05}.

A pure spin magnetism results in a temperature-independent $\mu_{\rm eff}= 1.73$ $\mu_{\rm B}$ \cite{Hat05}, as it is found for example of $S$ = 1/2 Cu$^{2+}$ which has a 3$d^9$ electron configuration. It is however the case that there is a quenching of orbital contribution to $\mu_{\rm eff}$ with the reduced symmetry, e.g. $C_{2\nu}$, so that spin-orbit coupling has no effect. The effective magnetic moment is 0.81 $\mu_{\rm eff}$ per Ag atom of the hydrogen reduced cluster (see Table \ref{tab:Magnetization}). If it is compared to the pure spin effective moment, only $<1$ spin is shared among two Ag atoms or about $\sim3$ unpaired spins are per nominal Ag$^+_6$ cluster. It is merely assuming pure $S=1/2$ state, therefore suggesting that there is a fraction of EPR silent high-spin states with $S>1/2$, which in turn would mean that a significant fraction would have to be diamagnetic ($>45\%$). Thus, while the exact composition may vary from one sample to another it is clear that the results can only be explained with a mixture of different magnetic states such as diamagnetic, EPR active spin $S=1/2$ and EPR silent high-spin species.

It is conceivable that a contribution of the orbital angular momentum is responsible for a remarkably high $g$ factor ($g_{\rm iso}\approx2.028$) of the EPR active species. An enhanced diamagnetism has been predicted for small metal clusters and was ascribed to the larger radius of the ring currents in the overall cluster molecular orbitals compared with the atomic orbitals which are normally held responsible for diamagnetism \cite{Kre88}. It will depend also on the number of conduction electrons which is available in the cluster, and thus on the cluster charge. In terms of charge, a diffuse of electron density renders a deficiency of the electron inside the quantum sphere of the Ag clusters leading to a hole contribution to the positive shift of $g$ tensor.

It is assumed that 5$s$ electrons are mainly responsible for spin paramagnetism of reduced Ag$^+_6$ clusters. Additional unpaired electrons are abstracted from hydrogen molecules to the cluster during the reduction. It implies that the more empty orbital are needed for further occupations of $s$ electrons with parallel spins. Therefore, the allowed electronic energy states spread out into the conduction band and the orbital degeneracy can be removed by spin-orbit coupling. Since conduction band is a band of orbitals, which are high in energy, temperature increases basically lead to a continuous depletion of the number of electrons spin-up, thus reducing magnetization of the system.

Tight-binding molecular dynamics study demonstrated that HOMO of the Ag clusters are located in the $s$-like electronic states \cite{Zha01}. TDDFT calculations revealed that the excitations of $5s$ electrons from the doubly degenerate HOMO to LUMO are responsible for the absorption spectra of Ag$_6$ clusters with $C_{5 \nu}$ symmetry \cite{Zha06}. This is because HOMO consists of 77.1\% $5s$, 20.6\% $5p$, and 4.8\% $4d$ orbitals. However, an effect of $d$ orbital is not negligible, there is an contribution to the absorption spectra. It is obvious that the total susceptibility contains a diamagnetic contribution from the $4d$ electrons. Unfortunately, it is impossible to distinguish diamagnetism, since the subtraction/correction of a temperature-independent $\chi_0$ term for magnetization excludes a determination of any diamagnetic or a temperature-independent paramagnetic state.

\begin{table}[ht]
\small
\caption{Magnetic moments and fit parameters derived from experimental $\chi(T)$ data for different reduction conditions. Measurements as a function of temperature were performed at a fixed magnetic field of 2000 Oe. Note that the unit of the magnetic field is given as Oe in CGS, since the unit of $\mu_{\rm eff}$ is not different in both SI and CGS system.}
\label{tab:Mag_val}
\begin{tabular*}{0.475\textwidth}{@{\extracolsep{\fill}}lcccccc}
\hline
Samples 	 & $\mu_{\rm eff}$  & $f$   & $\mu_{\rm cl}$  & $\chi_0(\times10^{-6})$ & $C_{\rm m}(\times10^{-5})$ & $\theta$\\
             & $(\mu_{\rm B})$  & --    & $(\mu_{\rm B})$ & (emu/mol)               & (emu$\cdot$K/mol)          & (K)     \\
\hline
D$_2$ red    & 0.13             &0.015  & 2.6             & -17                     & 16.1                       & -0.6(1) \\
H$_2$ des    & 0.09             &0.006  & 2.9             & +23                     & 9.8                        & -0.5(1) \\
H$_2$ red    & 0.07             &0.004  & 3.1             & -12                     & 6.1                        & -0.2(2) \\
\hline
\end{tabular*}
\end{table}

A temperature dependence of magnetic susceptibility $\chi(T)$ was also measured at the different magnetic fields of 1000, 3000 and 10000 Oe, respectively. All reduced samples were sealed in the NMR tubes prior to measurements. The fit of $\chi(T)$ was severely hampered by the presence of antiferromagnetic transition which was observed in a lower temperature range of $T<$ 6 - 13 K. An average of the Curie temperatures was close to $|\theta|\approx-$ 9 K. Above 13 K an antiferromagnetic state assumes paramagnetic exhibiting Curie-Weiss behavior to contribute to magnetic susceptibility. By fitting data, the highest moment was in the order of $\mu_{\rm eff}\approx0.11$ $\mu_{\rm B}$ per Ag atom in the deuterium-reduced cluster at 3000 Oe, whereas the lowest was $\mu_{\rm eff}\approx0.095$ $\mu_{\rm B}$ for the re-reduced cluster at 10000 Oe. The values of $\mu_{\rm eff}$ are generally quite close to each other in spite of being measured at 3000 and 10000 Oe. In contrast, with regard to interatomic spin-spin couplings in the range of cm$^{-1}$, one should consider that an external magnetic field of 10000 Oe corresponds to the energy equivalent wavenumber of 0.466864 cm$^{-1}$, so that spin-spin coupling and applied field have to be regarded as competing effects \cite{Hat05}.

A temperature dependence of magnetic susceptibility $\chi(T)$ generally survived at 1000 Oe. However, experimental data points were severely scattered since the magnetic field strength was too weak to align the magnetic moments sufficiently, while a temperature increased further. The spin-system entropy is a function of the population distribution, hence the spin entropy is a function only of an external magnetic field - temperature ratio ($H/T$). The entropy is lowered by the field \cite{Kit05}.

\section{Conclusion}

EXAFS analysis yields high precision values for the local environment of {\it atoms} in silver nano-clusters but not a complete structure and symmetry. Atomic arrangements in clusters are well ordered locally, but are not long-range ordered as was expected. Precisely, {\AA}ngstr\"om resolution distance data are experimentally available from the atomic pair (Ag-Ag) in clusters under different reduction conditions. The inter-atomic Ag-Ag distance is contracted on the order of $\approx1/10$ (9.5\%) of an {\AA}ngstr\"om as compared to the bulk 2.889 {\AA} distance, and this change is sufficient to render a size effect on cluster properties. The number of neighbors in the first and second coordination shell cannot provide information on the cluster geometry, but is used to estimate the mean size. Since silver clusters are supported systems, the interaction with the matrix can generally influence symmetry and geometry of the nano-clusters in terms of energy and again the analysis of the first two shells is not sufficient to determine the cluster shape.

The bulk Ag is typically diamagnetic, but exhibits paramagnetism when atoms are dispersed into smaller clusters in the support pores. Odd numbers of electrons are distributed in singly occupied molecular orbitals (usually HOMO) which strongly delocalizes over all atoms of the nominal Ag$^+_6$ cluster. And the conduction electron band becomes indeed discrete, and energy level separation is to be comparable to $\mu_{\rm B}H$. An odd-electron susceptibility leads to a Curie-Weiss dependence on temperature fluctuations, and magnetization is dominated by paramagnetic alignment of the spins at the lowest temperature of 1.8 K. Only 0.04\% of the exchanged atoms contribute, forming a fraction of magnetically active clusters, suggesting that a mixture of diamagnetic, EPR active spin $S=1/2$ and EPR silent high-spin species is present.

A single set of the numerical spectrum simulation for electron spin Hamiltonian parameters strongly supports that a single, well-defined paramagnetic silver cluster structure dominates the distribution of non-uniform, magnetically inactive cluster species. $g>g_{\rm e}$ indicates a positive charge on the cluster. The structural ordering of the Ag$^+_6$ cluster is a close-packing (even atoms) with the electronic open shell for a doublet ground state and the electronic {\it Zeeman} splitting is isotropic for symmetry reasons for the unpaired $5s$ electrons. The isotropic coupling constant ($A_{\rm iso}$) of the Ag$^+_6$ cluster is over 10 times smaller than the isotropic hyperfine coupling of the isolated Ag atom in the gas phase. No hyperfine anisotropy due to $p$ or $d$-orbital contributions is observed, and the unpaired spin is uniformly distributed among the silver nuclei rather than localized on a single Ag atom.

\section*{Acknowledgement}

The author is thankful to Prof. E. Roduner for his valuable discussions. The author is grateful to Prof. J. van Slageren and Prof. J. van Bokhoven for their valuable discussions on SQUID and X-ray absorption spectroscopy results. A.B. was supported by the Deutsche Forschungsgemeinschaft for a doctoral thesis scholarship through the Research Training Group-448 "Advanced Magnetic Resonance Type Methods in Materials Science" at the University of Stuttgart.

\end{document}